# The Reversible Spin Switch by External Control of Interval Distance of CuPc and $C_{59}N$ with the investigation of DFT

Min Wang, Yan Zhou, Sui Kong Hark, Xi Zhu

School of Science and Engineering,
The Chinese University of Hong Kong,Shenzhen,
Shenzhen, Guangdong, 518172

**Abstract**:  In this paper, we introduce a new kind of spin switch based on a joint system of copper phthalocyanine (CuPc) and $C_{59}N$. Using density functional theory, we investigate the total magnetic moment of this system when gradually changing the interval distance between two molecules. The spin hopping happens during the critical distance with very low energy. This phenomenon shows a possible reality of reversible spin switch by external control of the interval distance. With orbital analysis and electron transfer consideration, the form of $C_{59}N^{+}$- $CuPc^{-}$ ion pair support this spin hopping phenomenon.

## 1. Introduction

The traditional spin switch materials candidates are mainly contains the superconducting metal layers sandwiched between two ferromagnetic materials [1; 2]. A change in the electric potential of the electrode causes the magnetic semiconductor to make a reversible transition between a ferromagnetic state and a paramagnetic state. While in the molecular scale, this method for spin switch meets some problems. First, it is difficult in the experiment to reduce the thickness of the metal layers in the several atoms level and the ferromagnetic or anti-ferromagnetic property may not be kept the same as that in macroscale. Furthermore, since the size reduced the periodic condition is in doubt, the external voltage may push both the valence and conductor bands in the opposite direction and make the system become insulator. Therefore, it is necessary to develop new materials or device for the molecular spin switch for the further electronic application. The spin crossover phenomenon in molecular inorganic compounds is one of the most spectacular phenomena leading to a switching between the high-spin ("1") and the low-spin ("0") states of the molecule by several means such as temperature[3], pressure[4], and light [5]. Now one new method could be applied for spin switch device due to a very recent molecular push-button switch experiments reported by Kröger *et al* [6]. In Ref. [6], a tin ion was pulled ("1") and pushed ("0") through a phthalocyanine (Pc) compound by an STM tip in an absorbed self-assembled layer on a substrate surface. Encouraged by this experiment and the needs for new type spintronics devices, herein we perform density functional theory to provide a candidate reversible spin switch by the external control of interval distance between two molecules, CuPc and $C_{59}N$.

The phthalocyanine, one kind of organic semiconductors, plays an essential role in electronic and optical devices such as organic lighting emitting diodes [7; 8], organic field effect transistors [9; 10], gas sensors [11] and organic solar cells [12; 13] as well. The prospect of switching the spin in the metal phthalocyanine ring is a particularly interesting one, as this could be used, for example, for spin-dependent electric transport through biomolecular devices. Previously, the structural properties of phthalocyanine molecules in bulk and thin films have been studied by various techniques, such as ultraviolet photoelectron spectroscopy (UPS) [14], X-ray diffraction (XRD) [15] and X-ray photoelectron spectroscopy (XPS) [16]. Also, electronic structure calculations have been carried out for molecules [17].

$C_{59}N$, the one ball nitrogen doped modification of fullerenes, has a rich chemistry due to its enhanced reactivity as compared to pristine fullerenes and can be synthesized in macroscopic amounts chemically [18]. The electronic state of $C_{59}N$ and its derivatives is strongly modified compared to $C_{60}$ [19]. Spintronics properties of $C_{59}N$, such as the experiments of a molecular rectifier based on a single $C_{59}N$ molecule in a double-barrier tunnel junction was reported as well [20]. In this work, it is the first time to consider these two molecules - CuPc and $C_{59}N$ together as a spintronics system, and expend our methodology in this area.

## 2.  Results and discussion

Here we consider such case - gradually moving the $C_{59}N$ molecule towards the CuPc molecule which has been fixed. And then simulate the total spin magnetic moment of the system at each step. All the calculations are based on density functional theory [21; 22], as implemented in OpenMx code [23-25], with the local spin density approximation (LSDA) to the exchange-correlation potential. And the cut-off energy and electronic temperature are chosen as 140.0 $Ry$ and 300 $K$ respectively. As the numerical atomic orbitals (PAOs), we use C4.0-s2p2, N4.5-s2p2, H4.0-s1, and Cu4.5-s2p2d2 for C, N, H and Cu respectively. The abbreviation of each basis orbital means as follow: atomic symbol Cu, then its cutoff radius is 4.5 (in atomic units), and next s2p2d2 means that each two primitive orbitals are employed for each of s, p and d orbitals respectively. We have also used the LSDA+$U$ approach, in which an additional on-site Hubbard-$U$ term is included on the copper atom. From LSDA+$U$ calculations, for free CuPc molecule are similar to the one studied here. In the case of $C_{59}N$ molecule, because of the unpaired electron, the total spin magnetic moment of $C_{59}N$ is calculated as 0.91 $\mu B$ and that of CuPc is about 1 $\mu B$. And we also found that the calculated magnetic moment of Cu atom in the single CuPc molecule is 0.52 $\mu B$, which is close to the 0.50 $\mu B$ in CuPc dimer [26].

Here we consider the joint structure of the CuPc / $C_{59}N$. When the $C_{59}N$ is moving towards to the CuPc gradually, we study the total magnetic moment of the system with changing the interval distance of the two molecules as shown in the insert of Fig. 1. At each step, first change the distance between two molecules then simulate the total magnetic moment of the system; the step of distance changing is 0.1 $\mathring{A}$. When the distance is equal to or larger than 2.6 $\mathring{A}$, the total spin moment is larger than 1.6 $\mu B$ (see Fig. 1). But only after decreasing 0.1 $\mathring{A}$ of the distance, the total magnetic moment drops down, and stays as a level about 0.13 $\mu B$ (see Fig. 1 and Table 1). Furthermore, we calculate the energy of the system at A and B states, and the energy change between two states is only about 0.136 *eV*. This little energy gap shows a possibility of a realization for the distance control of spin switch device: decreasing or increasing 0.1 *Å*, then the total spin hopping.

As a comparison, in Ref. [6] the distance for the metal ion protruding or sinking is less than 0.1 *nm*, while the calculated energy barrier is larger than 1.5 *eV*, which needs to decrease for further applications. However, for our structure it has much smaller changeable energy barrier and much smaller distance between two states. It is also possible to act as a "traditional button" to give different spin signals: high-spin ("1") and low-spin ("0"), when pull ("1") and push ("0") certain molecule to change interval distance between CuPc and $C_{59}N$.

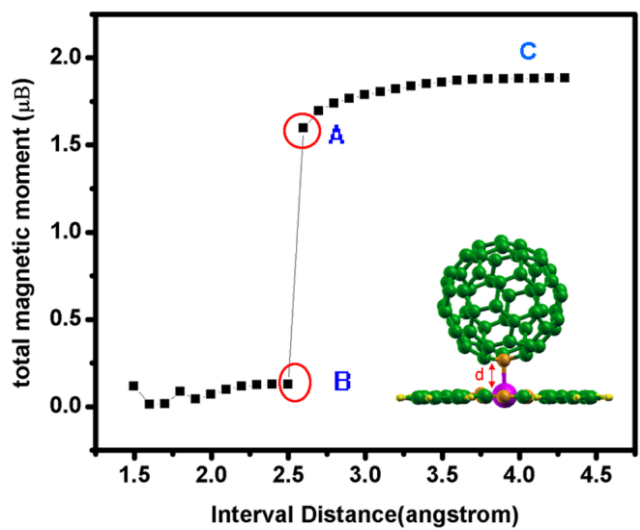

Fig. 1. Total magnetic moment of the CuPc / $C_{59}N$ system is changing with the interval distance. The critical status before and after the magnetic moment hopping is marked as A and B, the interval distance for A and B are 2.5 *Å* and 2.6 *Å* respectively. And at the right bottom part, the schematic diagram points out the interval distance between two molecules.

Table 1. Total magnetic moment of CuPc / $C_{59}N$ system v.s. interval distance.

| Interval Distance | Total magnetic | Interval Distance | Total magnetic |
|---|---|---|---|

| (Å) | moment ($\mu B$) | (Å) | moment ($\mu B$) |
|---|---|---|---|
| 1.5 | 0.12 | 3.0 | 1.79 |
| 1.6 | 0.01 | 3.1 | 1.80 |
| 1.7 | 0.02 | 3.2 | 1.82 |
| 1.8 | 0.08 | 3.3 | 1.84 |
| 1.9 | 0.04 | 3.4 | 1.85 |
| 2.0 | 0.07 | 3.5 | 1.86 |
| 2.1 | 0.10 | 3.6 | 1.87 |
| 2.2 | 0.12 | 3.7 | 1.87 |
| 2.3 | 0.12 | 3.8 | 1.88 |
| 2.4 | 0.13 | 3.9 | 1.88 |
| 2.5 | 0.13 | 4.0 | 1.88 |
| 2.6 | 1.60 | 4.1 | 1.88 |
| 2.7 | 1.70 | 4.2 | 1.88 |
| 2.8 | 1.74 | 4.3 | 1.88 |
| 2.9 | 1.77 | | |

In order to further understand the spin hopping in the system, we analyze the orbitals and list the HOMOs of state A, B and C in Fig. 2. In state A, there is weak interaction between the two molecules, since the HOMOs of the two molecules are both half occupied and the two half-occupied orbitals are apt to combine together to low the total energy. Before reaching the whole combination, the two half-occupied orbitals will partially overlap as the HOMO of $C_{59}N$ is -3.13 $eV$, and that of CuPc is -3.45 $eV$. For the CuPc, the $3d$ orbital $b_{1g}$, which is half occupied, is positioned in the gap between the Pc HOMO and LUMO [27]. In Fig. 2 (a) and (c), we can see before reaching state A, when scaling the interval distance of CuPc and $C_{59}N$, the system lowers the energy and the magnetic moment, reduces the volume of the HOMO of $C_{59}N$, and increases the amount of charge transport from $C_{59}N$ to CuPc. And at state B, without the orbital degenerating, the $C_{59}N$ totally loses the $3s$ electron to fill CuPc's half-occupied $dx^2$-$y^2$ orbital full and forms the $C_{59}N^+$ - $CuPc^-$ ion pair. As shown in Fig. 2 (b), the HOMO of the system is the Cu $dx^2$-$y^2$ orbital. When the CuPc is reduced, all the electron deposition of electrons into the Pc $2e_g$ orbital, bypassing the lower $b_{1g}$ ($dx^2$-$y^2$) [27], then it shows that the shape is the same as the isolated CuPc molecule shown in Fig. 2 (d).

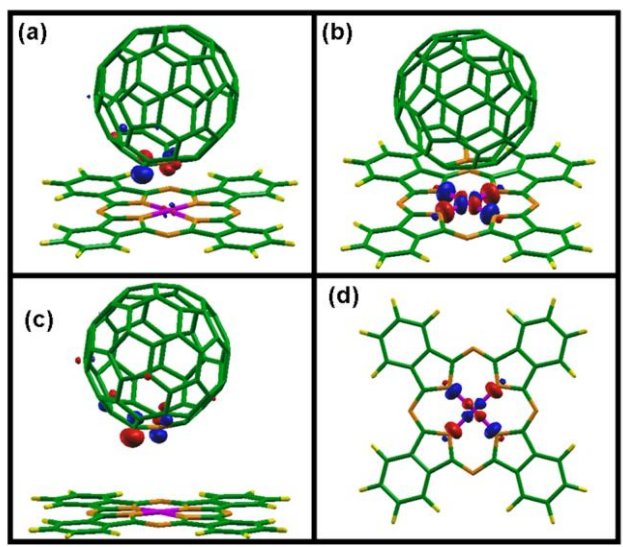

Fig. 2. (a), (b), (c) and (d) are the HOMOs of state A, B, C (the states shown in Fig. 1) and single CuPc respectively. Blue and red lobes represent positive and negative phases, respectively.

Moreover, with the consideration of electron transferring between two molecules shown in

Fig. 3, single electron occupies on each HOMO orbital represented by arrows. Since the energy of CuPc's HOMO is lower than that of $C_{59}N$, when the two molecules are closer, the electron would transfer from HOMO of $C_{59}N$ to that of CuPc, which will cause CuPc$^-$ and leave a hole to form $C_{59}N^+$. Then generate $C_{59}N^+$ - CuPc$^-$ ion pair to low the energy. Therefore, during the spin hopping, the energy change is very small.

When continuing short the interval distance, the low-spin state remains within 0.13 $\mu B$

but the total energy increases fast with few application potentials for huge exotic energy.

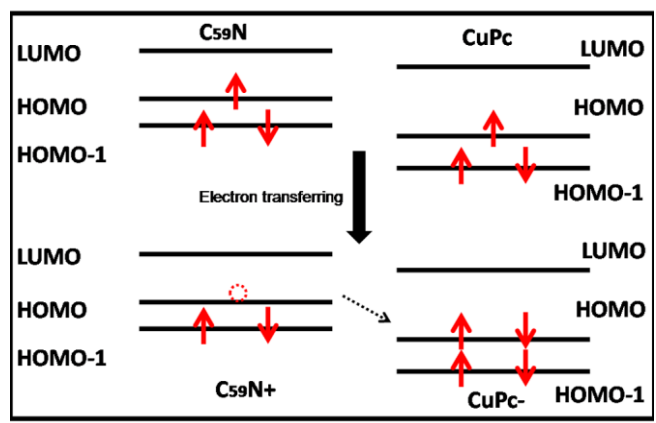

Fig. 3. Orbital energy levels for the $C_{59}N$ and CuPc molecules. And the upper diagram shows two molecules' orbitals and bottom schematic diagram shows electron transfers from HOMO of $C_{59}N$ to that of CuPc. The arrows and ring represent electrons and holes, respectively.

## 3. Conclusion

In summary we have shown that in the joint structure of CuPc / $C_{59}N$, the spin hopping can happen when decreasing the interval distance between two molecules, which can be a possible reality of spin switch controlled by interval distance. And we also analyze HOMO orbitals of related states for further understanding of spin hopping. With the consideration of electron transfer, the form of $C_{59}N^{+}$- $CuPc^{-}$ ion pair applies an additional support for the spin hopping.

**Acknowledgement**

This work is financially supported by xxx.

**Figure captions**

Fig. 4. Total magnetic moment of the CuPc / $C_{59}N$ system is changing with the interval distance. The critical status before and after the magnetic moment hopping is marked as A and B, the interval distance for A and B are 2.5 $\mathring{A}$ and 2.6 $\mathring{A}$ respectively. And at the right bottom part, the schematic diagram points out the interval distance between two molecules.

Fig. 5. (a), (b), (c) and (d) are the HOMOs of state A, B, C (the states shown in Fig. 1) and single CuPc respectively. Blue and red lobes represent positive and negative phases, respectively.

Fig. 6. Orbital energy levels for the $C_{59}N$ and CuPc molecules. And the upper diagram shows two molecules' orbitals and bottom schematic diagram shows electron transfers from HOMO of $C_{59}N$ to that of CuPc. The arrows and ring represent electrons and holes, respectively.

**Table captions**

Table 2. Total magnetic moment of CuPc / $C_{59}N$ system v.s. interval distance.